\documentstyle[11pt]{article}

\small
\begin{document}

\title{STRANGENESS CONTENT OF THE NUCLEON
WITHIN THE NJL SOLITON MODEL\footnote{\noindent Supported by the Deutsche
Forschungsgemeinschaft (DFG) under contract number Re--856/2-2.}
\footnote{\noindent Talk given at the conference {\it Hadron 95},
July 9th -14th, 1995, Manchester.}}

\author{A. ABADA,
H. WEIGEL\thanks{\noindent Supported by a Habilitanden scholarship
of the DFG.}, R. ALKOFER, H. REINHARDT \\
Institute for Theoretical Physics, T\"ubingen University, \\
Auf der Morgenstelle 14, D-72076 T\"ubingen, Germany}

\date{}
\maketitle

\begin{abstract}
We investigated the form factors of the nucleon associated with the nonet
vector current in the framework of the generalized Yabu--Ando approach to the
chiral soliton of the Nambu--Jona--Lasinio model. We introduce a variational
parameter to improve the predictions for the baryon mass splittings and,
in addition, we take into account $1/N_c$ rotational corrections to
accommodate the empirical isovector magnetic moment of the nucleon.
We find that the strange magnetic moment lies between -0.05 and 0.25.
\end{abstract}

In the near future some experiments, such as the SAMPLE experiment at
MIT/Bates~\cite{McK89} and the $G^0$ experiment at CEBAF~\cite{Bec91}
seek to extract information on the strange magnetic moment $\mu_s$ of the
nucleon. Since up to now no experimental data are available for $\mu_s$
 as well as for the strange electric vector form factor,
theoretical predictions are helpful to obtain some suggestions for these
quantities. A first estimate of
the matrix element of the strange vector current between nucleon
states was carried out in reference~\cite{Ja89} by performing a three--pole
vector meson fit to dispersion relations~\cite{Ho74}. This yielded
$\mu_s=-0.31$.
Later on the matrix element $\langle N|{\bar s}\gamma_\mu s|N\rangle$ has been
studied in the Skyrme model~\cite{Pa91}
and the Skyrme model
with vector mesons~\cite{Pa92} with the results -0.13  and -0.05 for $\mu_s$,
respectively. In a simple constituent quark model, Ito~\cite{Ito} obtained
$\mu_s=-0.125$.
For further details on the status of related studies we refer to the review
article by Musolf {\it et al}~\cite{Mu94}. In this communication we
present the prediction of the Nambu--Jona--Lasinio (NJL)~\cite{NJL} model
for the strange form factors of the nucleon.

The NJL model has the advantage of imitating the quark flavor dynamics of QCD
at low energies and containing both reference to explicit quark degrees
of freedom as well as the fruitful concepts of chiral symmetry and its
spontaneous breaking. In addition, after bosonization~\cite{ER}, i.e.,
integrating out the quark fields in favor of the meson
fields \footnote{\noindent In this work we have considered only scalar and
pseudoscalar mesons.}, the NJL model contains soliton solutions~\cite{Al94}.
By using the generalized Yabu--Ando approach~\cite{YA} to these solitons
one is able to construct baryon wave functions and to investigate
the effects of the strange quarks. Here we will concentrate on the
matrix elements of the strange vector current ${\bar s}\gamma_\mu s$
between nucleon states.
The details of these calculations are given in reference~\cite{WAAR}.
Here we only like to mention  that
we have ignored recoil corrections which means that our results are only valid
near zero momentum transfer.

The bosonized NJL model contains four independent parameters: the up (down)
and strange current masses, the dimensionful NJL coupling constant and the
cut-off \footnote{The NJL model is non-renormalizable and a
regularization scheme has to be specified. We adopt here the Schwinger's
proper time prescription which imply the introduction of a $O(4)$ invariant
cut-off $\Lambda$~\cite{ER}.}.
These have been fixed by fitting to low energy meson data: the pion mass
(135 MeV), the pion decay constant (93 MeV), and the kaon mass (495 MeV).
By virtue of the gap (Dyson--Schwinger) equation the remaining parameter
can be expressed in terms of the up-quark constituent mass~\cite{Al94}.
The NJL soliton model suffers from the too small predictions
for the baryon mass splittings and the isovector magnetic moment $\mu_V$
of the nucleon. This can be traced back
to the too small spatial extension of the self--consistent NJL soliton.
This problem can be circumvented by explicitly
including the $\rho, \omega,..$ mesons into the NJL model,
or by increasing by hand (with a variational parameter $\lambda$)
the spatial extension of the soliton.
%Although being more realistic, the first
%alternative is not feasible at present. For this reason
In this work we opt for the
second alternative to improve our predictions. We fixed this parameter
by optimizing the baryon mass differences with the result $\lambda=0.75$.
As a consequence the symmetry breaking is effectively increased, lowering the
(virtual) ${\bar s}s$ pairs in the nucleon. We have also estimated the
 $1/N_C$ rotational corrections (leading to a value for $\mu_V$ closer
to its experimental value ). We find that these $1/N_c$ corrections
also cause $\mu_S$ to decrease. Hence we consider the result
obtained from the self--consistent soliton ($\lambda=1$) as an upper bound.
Assuming that all corrections add coherently $\mu_S$ may even assume small
negative values. We therefore arrive at the NJL model estimate
\begin{eqnarray}
-0.05\le\mu_S\le0.25.
\label{estmus}
\end{eqnarray}
The lower bound corresponds to the case where $\lambda=1$ and
$1/N_c$ corrections are estimated to accommodate the empirical isovector
magnetic moment.
We have furthermore observed that the valence quark provides the major
contribution to $\mu_S$ although the vacuum part adds coherently to the
strange magnetic form factor.
For the strange electric form factor of the nucleon, $G_{E,S}(q^2)$,
 the valence quark and vacuum contributions almost cancel each
other causing $G_{E,S}(q^2)$ to barely deviate from zero.
The NJL model estimate for the slope of this form factor
is
\begin{eqnarray}
-0.25{\rm fm}^2\le\langle r_S^2\rangle\le-0.15{\rm fm}^2.
\label{estrs2}
\end{eqnarray}
This result compares well with that of the Skyrme model~\cite{Pa91}.
As $G_{E,S}(q^2)\approx 0$ it is easy to understand that small effects may
modify this result. As an example we refer to the $\phi-\omega$ mixing,
which is neither included in the present model nor in the Skyrme
model~\cite{Pa91,Co93}.

\end{document}